\documentclass{emulateapj}
\usepackage{epstopdf}
\usepackage{graphicx}

\shorttitle{HI Content in Groups of Galaxies}
\shortauthors{Odekon et al.}
\begin{document}

\title{The HI Content of Galaxies in Groups and Clusters as Measured by ALFALFA}
\author{Mary Crone Odekon}
\affil{Department of Physics, Skidmore College, Saratoga Springs, NY 12866, USA; mcrone@skidmore.edu}
\author{Rebecca A. Koopmann}
\affil{Department of Physics and Astronomy, Union College, Schenectady, NY 12308, USA; koopmanr@union.edu}
\author{Martha P. Haynes}
\affil{Cornell Center for Astrophysics and Planetary Science, Cornell University, Ithaca, NY 14853, USA; haynes@astro.cornell.edu}
\author{Rose A. Finn}
\affil{Physics Department, Siena College, Loudonville, NY 12211, USA; rfinn@siena.edu}
\author{Christopher McGowan}
\affil{Department of Physics, Skidmore College, Saratoga Springs, NY 12866, USA; cmcgowan@skidmore.edu}
\author{Adina Micula}
\affil{Department of Physics, Skidmore College, Saratoga Springs, NY 12866, USA; amicula@skidmore.edu}
\author{Lyle Reed}
\affil{Department of Physics, Skidmore College, Saratoga Springs, NY 12866, USA; lreed@skidmore.edu}
\author{Riccardo Giovanelli}
\affil{Cornell Center for Astrophysics and Planetary Science, Cornell University, Ithaca, NY 14853, USA; riccardo@astro.cornell.edu}
\author{Gregory Hallenbeck}
\affil{Department of Physics and Astronomy, Union College, Schenectady, NY 12308, USA; hallenbg@union.edu}  
\begin{abstract}
We present the HI content of galaxies in nearby groups and clusters as measured by the 70\% complete Arecibo Legacy Fast-ALFA (ALFALFA) survey, including constraints from ALFALFA detection limits. Our sample includes 22 systems at distances between 70-160 Mpc over the mass range $12.5<\log M/M_{\sun}<15.0$, for a total of 1986 late-type galaxies. 
We find that late-type galaxies in the centers of groups lack HI at fixed stellar mass relative to the regions surrounding them.  Larger groups show evidence of a stronger dependence of HI properties on environment, despite a similar dependence of color on environment at fixed stellar mass.  We compare several environment variables to determine which is the best predictor of galaxy properties; group-centric distance $r$ and $r/R_{200}$ are similarly effective predictors, while local density is slightly more effective and group size and halo mass are slightly less effective. 
While both central and satellite galaxies in the blue cloud exhibit a significant dependence of HI content on local density, only centrals show a strong dependence on stellar mass, and only satellites show a strong dependence on halo mass.  Finally, we see evidence that HI is deficient for blue cloud galaxies in denser environments even when both stellar mass and color are fixed.  This is consistent with a picture where HI is removed or destroyed, followed by reddening within the blue cloud. Our results support the existence of pre-processing in isolated groups, along with an additional rapid mechanism for gas removal within larger groups and clusters, perhaps ram-pressure stripping. 
\end{abstract}


\section{Introduction} 

The distribution of galaxy properties is bimodal in the parameter space of mass and star formation activity. In all but the largest galaxies, star formation tends to be either quenched and the stellar population red, or active and the stellar population blue. The lack of galaxies in between these two populations suggests that quenching mechanisms occur rapidly, or else that the nature of each galaxy is determined by events in the distant past.  

A clue to the nature of this bimodality is the strong environmental dependence of color, morphology, and other characteristics linked with star formation; quenching is clearly more pronounced in dense regions.  On the other hand, once galaxies are split into quenched and star-forming populations, the properties within each group are remarkably independent of environment at fixed luminosity or stellar mass. The environmental dependence is almost completely explained by the varying fraction of the two populations with environment.
This effect has been observed for color, shape, star formation rate, galaxy mass function, metallicity, and many other properties (e.g. Balogh et al. 2004, Blanton et al. 2005, Park et al. 2007, Mouhcine et al. 2007, Peng et al. 2010, van der burg et al. 2013, Hughes et al. 2013, Brough et al. 2013). 
Studies of environmental dependence thus bolster the picture that quenching is either rapid or depends on the distant past; if galaxies are gradually quenched as they enter dense environments, we should see a more obvious continuum of properties across regions of different density, not simply a continuous variation in the fraction of each population. 

Some quenching mechanisms, like ram-pressure stripping of gas in the dense inner parts of clusters, are indeed expected to occur rapidly, and their importance is supported by a variety of additional observations.  Muzzin et al. (2014), for example,
find that the distribution of post-starburst galaxies in phase space for a sample of $z\sim1$ clusters from the GCLASS survey implies quenching that is rapid and within the cluster virial radius.  
But quenched galaxies often appear at cluster-centric distances too large to be explained by the usual picture of ram-pressure stripping. Galaxies are apparently pre-processed in intermediate density environments before entering clusters, possibly through tidal interactions that remove gas or trigger star formation, or through the heating of cool gas in sufficiently large haloes.  For example, based on the SDSS NoSOCS sample, Lopes et al. (2014) conclude that pre-processing in groups is strong, with local density as the main driver for galaxy evolution.   

An additional type of quenched galaxies at large cluster-centric distances are ``backsplash" galaxies on orbits that have already taken them through a cluster center (e.g. Wetzle et al. 2014, Hirschmann et al. 2014). However, Bahe et al. (2013) conclude that even a combination of preprocessing in groups and backsplash is not sufficient to explain the quenching at high distances from cluster centers, pointing to ram pressure stripping along filaments far outside the virial radius of clusters. 

Observations of atomic hydrogen gas (HI) have the potential to reveal the first step of the quenching process, the removal or destruction of cool gas.  Observations of individual nearby galaxies indicate both tidal interactions in small groups (e.g. Yun et al. 1994) and stripping in clusters (e.g. Kenney et al. 2004), even at intermediate distances from cluster centers (Chung et al. 2007), while statistical studies of HI content in larger samples show that galaxies have less HI in high-density environments (e.g. Govanelli \& Haynes 1985, Solanes et al. 2001, Gavazzi et al. 2013, Denes et al. 2014), and that the degree of deficiency in these environments suggests ram-pressure stripping (Cortese et al. 2011). Indeed, the analysis of clustering of HI-detected galaxies as measured by the correlation function (e.g. Martin et al. 2012, Papastergis et al. 2013) shows that these galaxies are the least clustered galaxy population known.  

The degree to which galaxies are HI deficient in intermediate-density environments, and the range of galaxy masses and colors for which HI content is segregated, is more complicated (see Cortese et al. 2011).   
Using data from the HI Parkes All-Sky Survey, Sengupta \& Balasubramanyam (2006) find that galaxies in groups with diffuse X-ray emission have less HI than groups with no diffuse X-ray emission, but found no trend in HI content with X-ray luminosity among the X-ray detections.
Fabello et al. (2012) use a stacking technique to find average values (or upper limits) for galaxies without a sufficient 21-cm signal to be detected individually for a subset of the GALEX Arecibo SDSS survey of massive galaxies (GASS; Catinella et al. 2010).
Splitting galaxies into two bins of mass and four of density, they find that the less massive galaxies in particular have a stronger change in HI mass fraction with density than specific star formation does with density. 
The final data release from the GASS sample (Catinella et al. 2013) shows that massive galaxies in halos of mass $10^{13}-10^{14}$ have less HI than galaxies in lower-density environments.  
Similarly, using the 40\% complete catalog from the Arecibo Legacy Fast ALFA survey (ALFALFA; Haynes et al. 2011), Hess \& Wilcots (2013) find that HI detections are less concentrated than optically selected galaxies, with the fraction of HI detections decreasing with halo mass. 
Yoon \& Rosenberg (2015) find that while the HI detection fraction declines steadily in the center of haloes (especially massive haloes), galaxies detected in HI do not exhibit a significant variation in HI mass ratio as a function of group-centric distance within the virial radius. 

High-resolution radio observations of individual galaxies in intermediate-density environments show evidence of interactions that may be removing HI.  Denes et al. (2016) mapped the HI distribution for a sample of six HI-deficient galaxies in intermediate-density environments using the Australia Telescope Compact Array and found morphologies indicating ram-pressure stripping in at least two of them. Another suggestion that there is rapid quenching in groups is due to Walker et al. (2016), who find three galaxies in compact groups with mid-IR activity that is discrepant from what they expect based on HI content. Following the complementary technique of searching for \textit{intragroup} HI that might have been pulled or stripped from galaxies, Serra et al. (2015) find significant intragroup HI gas (10\% of the total) in IC 1459, and Borthakur et al. (2015) find faint gas in four Hickson Compact Groups that appears to be tidal in origin.

In this paper, we take the approach of comparing the HI content, at fixed galaxy type and stellar mass, in the centers of groups with the HI content of galaxies in a control region out to 4 Mpc surrounding each group. 
This process allows us to statistically test for pre-processing in intermediate-density, isolated groups relative to the regions immediately surrounding them -- a method that complements tests based on local density. 
For the first part of our analysis, we focus on a sample of eighteen groups, with a total of 1142 late-type galaxies, and use binning techniques as well as multiple linear regressions to examine galaxy HI content, adopting methods designed to include HI detection limits through survival analysis statistics.   
We then expand the sample to include four larger clusters, for a total of 1887 late-type galaxies, to test for the dependence of HI content in groups and clusters on several different environmental predictors, including group mass, local density, and halo central/satellite status.  
We also use this expanded sample to see whether HI content appears to be changing more quickly with environment than does galaxy color.

Our HI data are from the $\alpha.70$ catalog, the most recent data release from the ALFALFA survey (Giovanelli et al. 2005), which covers 70\% of the total survey area of 7000 deg$^{2}$ with an rms noise $\sim 2.2$ mJy and beam size $\sim 3.6$ arcminutes. 

We describe the selection of our sample in Section 2, the calculation of galaxy properties in Section 3, and our results in Section 4.  We summarize our conclusions in Section 5. Throughout this paper we assume the cosmological parameters $\Omega_{m}=0.3$, $\Omega_{\Lambda}=0.7$, and $h=0.7$.

\section{Sample and Galaxy Membership}

To construct the sample, we combined groups from the RASSCALs (Mahdavi et al. 2000) and MCXC (Piffaretti et al. 2011) catalogs.  
These catalogs include either X-ray measurements or upper limits, providing information on the amount of hot intragroup gas that could strip HI from galaxies),and they cover approximately the same part of the sky as our survey.
Optically-determined properties, including redshifts, were obtained from the Sloan Digital Sky Survey (SDSS; Ahn et al. 2014).
In order to have a control sample of galaxies surrounding each group, we include only groups for which a full 4 Mpc group-centric distance fits within both the ALFALFA and SDSS regions on the sky. (Note that there is significant overlap between the ALFALFA survey and the SDSS only in the northern galactic hemisphere.) Any group closer than 6 Mpc to a larger group, and overlapping with that group in redshift space, is  excluded, so that only the control sample regions can overlap; this happens with one pair of groups in our sample.  We also use the Yang DR7 catalog (see Yang et al. 2007) to describe the environment of each galaxy in our sample through halo membership and central/satellite status.

Most of our groups are too small or too poorly sampled to use caustics to define galaxy membership based on infall patterns.  Instead, we use the following method to systematically define a range in redshift space for each group.  We first consider the redshifts of all galaxies within a projected 3 Mpc radius of the group center. From the central redshift of the group, we go out in redshift space galaxy by galaxy until the velocity dispersion suddenly rises, indicating contamination by galaxies outside the group.  Specifically, if the change in the velocity dispersion is at least twice the average change for three galaxies in a row, the redshift is cut just before the first of these three galaxies.  To avoid noise due to small numbers, we start with the fifth galaxy out in redshift space.  The values for each of these parameters, including the initial use of a 3 Mpc projected radius, were varied and the final values chosen to maximize group membership while minimizing contamination by surrounding structures across the range of distances and environments for groups in our sample.

Table 1 summarizes the properties of our sample, which includes 22 groups and clusters approximately spanning the range $12<\log M/M_{\sun}<15$, with Coma at the high end. As shown in the table, we split the sample into four mass bins.  The two higher mass bins (a total of four clusters) comprise systems similar to those in previous studies of HI Deficiency in clusters (e.g. Solanes et al. 2001).  The two lower mass bins (a total of eighteen groups) form the sample we use to test for pre-processing in groups in Section 4.2.  We use the entire sample together, a total of 1887 late-type galaxies, to test for the dependence of HI content in groups and clusters on several different environmental predictors, including group mass, local density, and halo central/satellite status.

\begin{turnpage}
\begin{deluxetable*}{lllcrrrrrcclc}
\tabletypesize{\small}
\tablecolumns{13}
\tablewidth{0pc}
\tablecaption{Group Sample}
\tablenum{1}
\tablehead{ \colhead{}  & \colhead{\underline{Identifications}} & \colhead{} & \colhead{} &
\colhead{R.A.}    & \colhead{Dec}   & \colhead{cz\tablenotemark{b}}     & \colhead{Distance\tablenotemark{c}}  &
\colhead{$\sigma_p$\tablenotemark{d}} & \colhead{Log(M/M${_\sun}$)\tablenotemark{e}}   & \colhead{R$_{200}$\tablenotemark{f}}  & \colhead{N$_{g}$\tablenotemark{g}}& \colhead{Log(L$_X$/erg s$^{-1}$)\tablenotemark{h}} \\
\colhead{RASSCALs} & \colhead{MCXC}   & \colhead{Yang\tablenotemark{a}} 
& \colhead{Other} &\colhead{(J2000)} & \colhead{(J2000)} & \colhead{(km s$^{-1}$)} & \colhead{(Mpc)} &
\colhead{(km s$^{-1}$)} & \colhead{} & \colhead{(Mpc)} & \colhead{}& \colhead{} } 
\startdata
\sidehead{LogM/M$_{\sun}\sim$15.0}
  NRGb226 & J1259.7+2756 & 1 & Coma & 194.903 & 27.939& 6951 & 103  & 973& 14.9& 1.99 & 330 & 44.48 \\
\sidehead{LogM/M$_{\sun}\sim$14.5}  
  NRGs341&J1523.0+0836 &17& A2063 & 230.773&  8.603&10371 & 150 & 832& 14.7& 1.73 &99 & 44.06 \\
  NRGb155&J1144.6+1945 &3 & A1367 & 176.152& 19.759& 6442  &  97 & 758& 14.6& 1.58 &225 & 43.86 \\
         &J1516.7+0701 &24& A2052 & 229.183&  7.019&10438  & 152 & 625& 14.4& 1.32 &91 & 44.16 \\
\sidehead{LogM/M$_{\sun}\sim$14.0}
  NRGs110&J1100.8+1033&101& A1142 & 165.204& 10.560&10643  & 157& 580& 14.3& 1.23 &48 & 43.06 \\
  NRGs117&J1110.7+2842&7 & A1185  & 167.695& 28.706&  9601 & 141& 521& 14.2& 1.12 &148 & 42.98 \\
  NRGb049&             &575&       & 144.474& 17.053&  8445 & 125& 478& 14.1& 1.03 &43 & $<$43.3 \\
        & J1338.4+2644&3638&      & 204.603& 26.744&  8538 & 125& 454& 14.0& 0.98 &49 & 41.29 \\
        & J1440.6+0328&22 &  MKW 8 & 220.159&  3.476&  8051 & 118& 447& 14.0& 0.97 &105 & 43.28 \\
  NRGb206&        &304& WBL 404/408 & 186.011&  9.357&  7052 & 106& 436& 14.0& 0.95 &67 & $<$42.2 \\
  NRGb286 &            &1016 &      & 212.536&17.594 &  5062 &  74& 432& 13.9& 0.94 &45 & $<$42.1 \\
\sidehead{LogM/M$_{\sun}\sim$13.5}
  NRGb177&J1204.1+2020&15  &       & 181.048& 20.348& 7079 & 106& 421& 13.9& 0.92 &139 & 42.83 \\
  NRGs027&J0916.1+1736&119 &       & 139.029& 17.602& 8657 & 128& 415& 13.9& 0.91 &60 & 42.93 \\
  NRGs076&             &151 &WBL 251& 151.672& 14.430& 8915 & 132& 380& 13.8& 0.83 &51 & 42.65 \\
  NRGb247&J1329.5+1147&79 &  MKW 11& 202.384& 11.789& 6840 & 102& 366& 13.7& 0.81 &66 & 42.98 \\
  NRGb103&             &1511&       & 162.500& 16.121& 6453 &  97& 343& 13.7& 0.76 &36 & $<$42.2\\
  NRGb302&             &251  &      & 217.189& 11.369& 8019 & 118& 305& 13.5& 0.68 &48 & 42.36 \\
  NRGs317 &            &133  &      & 221.791& 13.706& 8938 & 131& 296& 13.5& 0.66 &54 & 42.82 \\
  NRGb104 &            &440 &       & 162.816&  8.645& 6493 &  98& 255& 13.3& 0.58 &44 & $<$42.6 \\
  NRGs015 &            &405 &       & 135.977&13.546 & 8626 & 127& 223& 13.1& 0.51 &44 & $<$42.3 \\
  NRGs353 &            &186  &       &235.476&28.284 & 9680 & 140& 140& 12.5& 0.33 &54 & $<$42.5 \\
  NRGb151 &            &499  &MKW 10 &175.539&10.275 & 6096 &  92& 124& 12.3& 0.30 &41 & $<$42.3 \\
\enddata
\tablenotetext{a}{Yang group identifications from the Model C catalog (Yang et al. 2007).}   
\tablenotetext{b}{Biweight mean of heliocentric velocities for galaxies iteratively determined to lie within $R_{200}$.}
\tablenotetext{c}{Group distance following the process used for the SFI++ catalog (Masters et al. 2006). Distances for groups with $cz>6000 km s^{-1}$ are from the biweight $cz$ corrected to the CMB frame divided by $H_{o}=70$, while distances for closer groups include flow model corrections for Virgo and the Great Attractor. Each galaxy is assigned the distance of its group.}
\tablenotetext{d}{Biweight velocity dispersion of heliocentric velocities from the SDSS for galaxies iteratively determined to lie within $R_{200}$.}
\tablenotetext{e}{Mass within $R_{200}$, estimated from the velocity dispersion according to Munari et al. (2013).}
\tablenotetext{f}{Radius at which the mean interior density is 200 times the critical density, estimated from the velocity dispersion according to Munari et al. (2013).}
\tablenotetext{g}{Number of blue cloud galaxies, identified as described in Section 3.3 of the text.}
\tablenotetext{h}{X-ray luminosity from the RASSCALs catalog or, if not in RASSCALs, the MCXC catalog. RASSCALS values are corrected to h=0.7.}
\end{deluxetable*}
\end{turnpage}
\section{Galaxy Properties}

Our galaxy data are drawn from the SDSS spectroscopic galaxy catalog and the 70\% complete ALFALFA galaxy catalog with optical counterparts, which includes photometric optical counterparts that are not in the SDSS spectroscopic catalog.  Combining these lists, about 5\% of galaxies yielded multiple position matches that upon inspection turned about to be different parts of the same galaxy.  In those cases, only the central SDSS object was retained.  Of the 3878 galaxies in the combined list, 128 were detected in HI but were not part of the SDSS spectroscopic catalog.

\subsection{Environment variables}
We consider several environment variables against which to compare HI content.  One is an in-group versus out-of-group (annular control region) distinction based on projected distance from the group center. 
We consider both a distinction based on a fixed radius of $r<1.5$ Mpc for in-group compared with 2.0 Mpc $<r<$ 4.0 Mpc for out-of-group and a distinction based on a normalized radius of $r/R_{200}<0.75$ compared with $1.0<r/R_{200}<2.0$. 

We also run regressions against group-centric distance $r$, normalized group-centric distance $r/R_{200}$, local density $\Sigma$, and halo membership status according to the Yang DR7 group catalog. To calculate local density, we find the number of galaxies per projected Mpc$^{2}$, averaged over the $2-10$ nearest neighbors.  Neighboring galaxies are limited to those within the redshift cut for each group, and to those brighter than $M_{r} = -19$ for completeness throughout the sample.  The range of nearest neighbors was chosen for its effectiveness in defining substructures for this magnitude cutoff and for the range of distances in our sample.

Figure 1 shows the environment in and around the group MKW~11.  The tendency for galaxies in the centers of groups to have less HI is evident in the left panel, where optically detected galaxies are more clustered than those with HI detections, consistent with Martin et al. (2012).  The control region extending out to 4 Mpc is clumpy and contains several Yang halos. 
This clumpy environment within the group and the surrounding control region is typical for the sample,   
and illustrates one of the ways that a comparison of groups with a surrounding control region complements a comparison based on density that is more localized. A particular issue is that if the groups in our sample are processing HI gas themselves, one might expect differences in HI content relative to the regions immediately surrounding them, not just differences relative to truly isolated galaxies.
(Note that our group selection process eliminates groups if their control regions contain clumps as large as the groups themselves, so it is meaningful to consider the group as the central object.)

\begin{figure*}
\vspace*{-25mm}
\epsscale{1.0}
\includegraphics*[width=500pt]{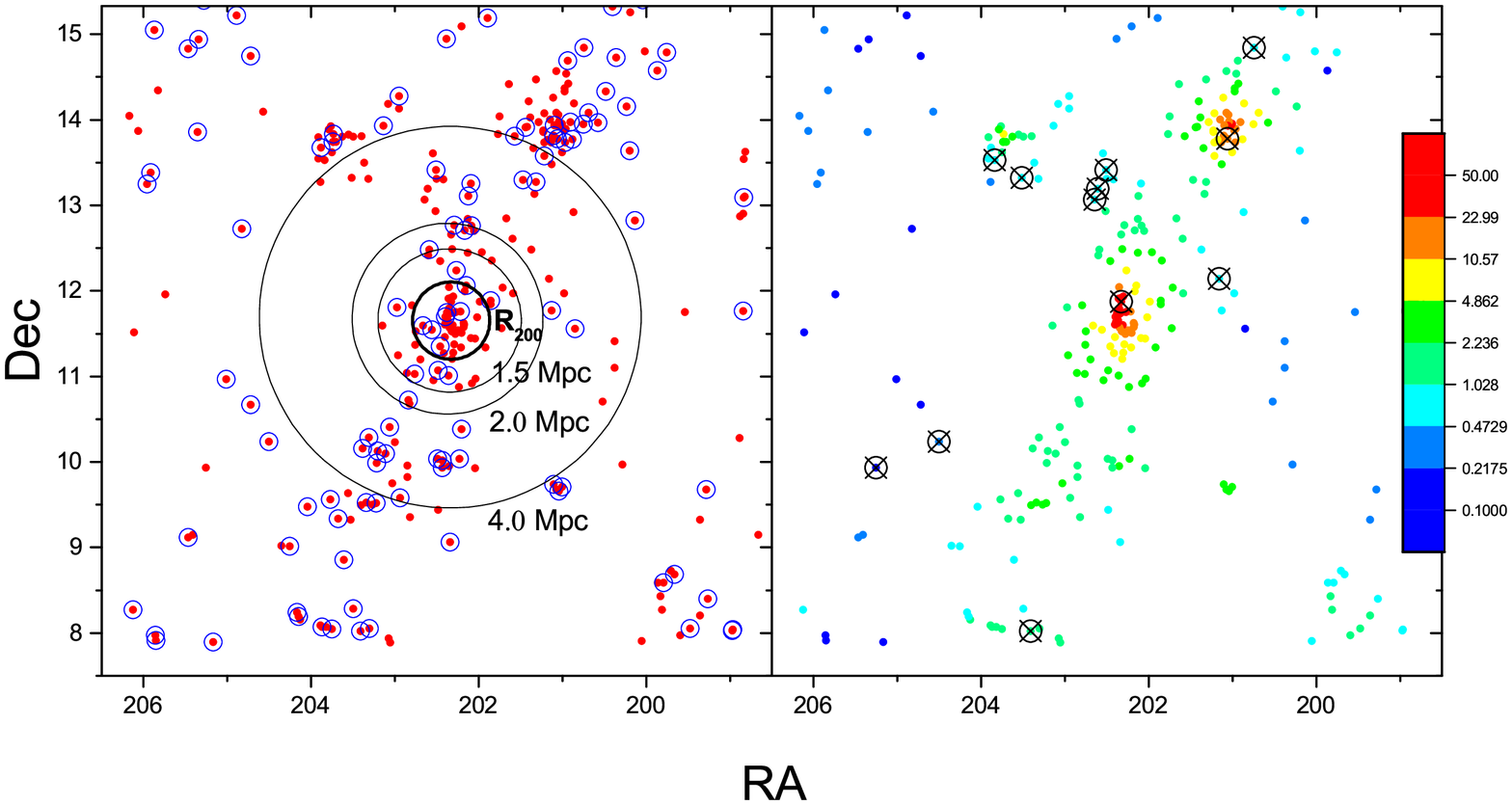}
\setlength{\abovecaptionskip}{-100pt} 
\caption{Distribution of galaxies in the region of MKW 11. In the left panel, red dots are optically selected galaxies and blue circles are ALFALFA detections.  The boundaries of the group region at 1.5 Mpc and control region from 2.0-4.0 Mpc are shown, as well as $R_{200}$.  In the right panel, color indicates local density, and black circled x's show the centers of Yang halos with log(M/M$_\sun)>12$.  Note that ALFALFA detections are less clustered than optically selected galaxies and that many Yang halos appear in low-density environments.}
\vspace{-10mm}
\end{figure*}
\begin{figure*}
\includegraphics*[width=500pt]{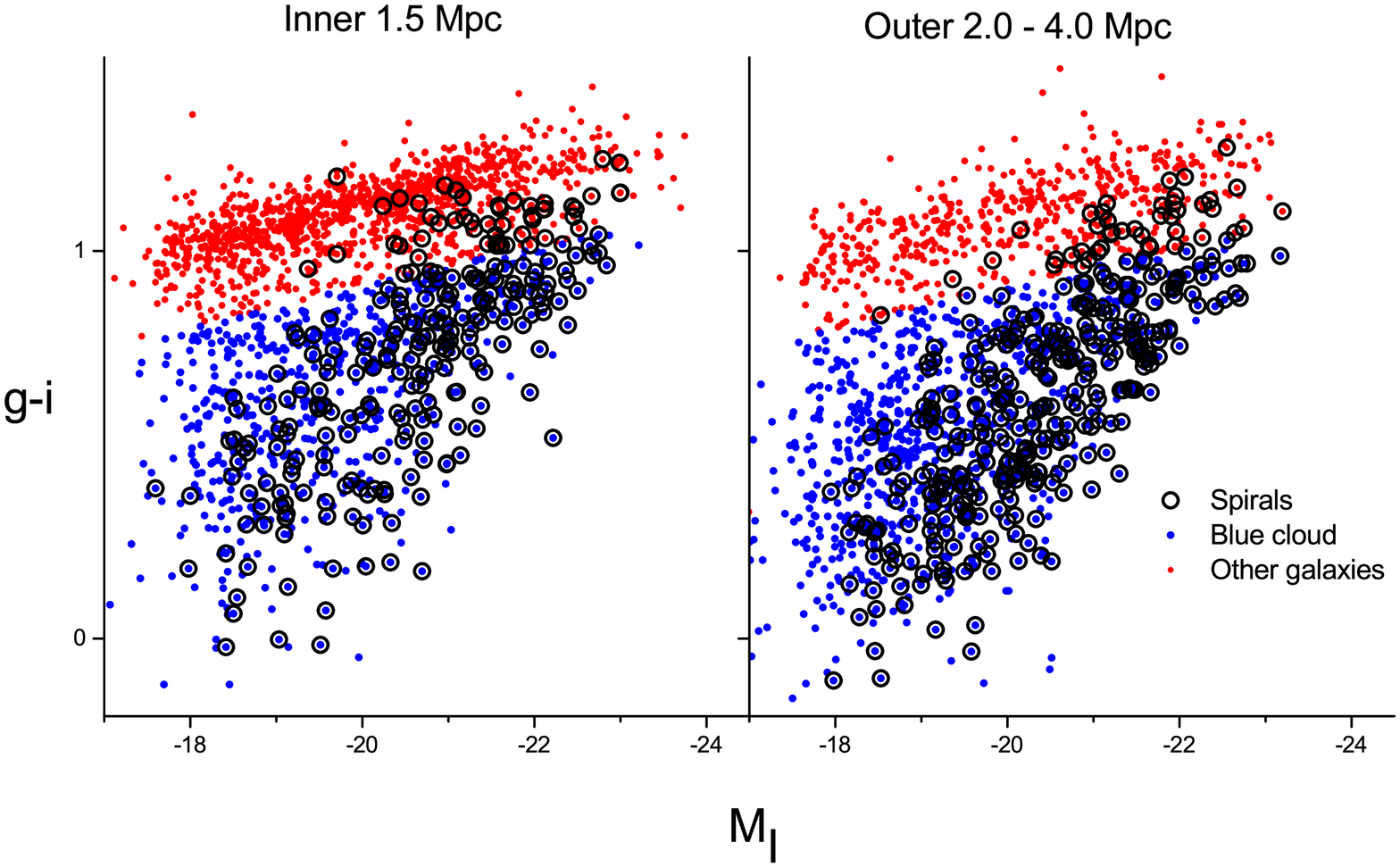}
\setlength{\abovecaptionskip}{-50pt} 
\setlength{\belowcaptionskip}{0pt} 
\caption{Color-magnitude diagrams for galaxies in the inner 1.5 Mpc and in the outer control sample, illustrating our selection of late-type galaxies.  Large circles show spiral galaxies according to the ``combined, clean" Galaxy Zoo criteria, blue dots show galaxies assigned to the blue cloud with 99\% confidence, and red dots show remaining galaxies.}
\end{figure*}

\subsection{Magnitudes, Masses, and HI Deficiency} 

Following the recommendations on the SDSS web site, we use SDSS cmodel mags for galaxy magnitudes and model mags for colors.  We use galactic extinction corrections from Schlafly \& Finkbeiner (2011) via the map tool at the NASA/IPAC Infrared Science Archive, and internal extinction corrections from Shao et al. (2007).  

We estimate stellar masses using Equation (7) in Taylor et al. (2011): $\log M_{\star}/L_{i}-0.68+0.70(g-i)$. We find good overall agreement between these masses and those from SED fitting models for the subset of galaxies in our sample where the SED fitting models are available on the SDSS web site.\footnote[1]{www.sdss.org/dr12/spectro/galaxy/}

In the analysis that follows we describe galaxy HI content three ways: the HI mass, the ratio of HI mass to stellar mass, and the HI Deficiency. HI masses are calculated using $M_{HI}/M_{\sun}=2.356\times 10^{5}D_{Mpc}^{2}S_{int}$, where $S_{int}$ is the integrated ALFALFA flux in janskys (Giovanelli et al. 2005).  
HI Deficiency is the logarithmic difference between the observed HI mass and the HI mass expected based on galaxy size:  $HIDef=\log M_{HI}^{exp}-\log M_{HI}^{obs}$ (Haynes et al. 1984).  Here we use the calibration of Toribio et al. (2011), who find that $M_{HI}$ correlates especially well with galaxy diameter according to the relationship $\log (M_{HI}/M_{\sun})=8.72+1.25\log D_{25}$. The r-band diameter $D_{25}$ in kiloparsecs is calculated from the size $isoA_{r}$ in the SDSS DR7, the galaxy inclination from the axial ratio $expAB$, and the number of kiloparsecs per arcsecond at the distance of the galaxy $adist$ according to $\log D_{25} = \log(2isoA_{r}0.39 \arcsec adist)+0.35\log(expAB)$. 

\subsection{Galaxy type}

While it is clear that HI detections are less clustered than optically selected galaxies, this observation is largely related to the well-known morphology-density relation for galaxies combined with the fact that late-type galaxies contain more HI gas than early-type galaxies.  In this paper we wish to address the more subtle question of whether HI content is different in galaxies when limited to late-type only.  It is particularly important to limit the sample by type if we wish to include upper limits for galaxies not detected in HI, many of which are early-type galaxies.  

We identify late-type galaxies in two ways.  One is by the Galaxy Zoo ``combined, clean" spiral criteria (Lintott et al. 2008), where at least 80\% of users classify a galaxy as right-hand spiral, left-hand spiral, or edge-on disk.  
The other is by color and magnitude: fitting all the 3878 galaxies to two two-dimensional Gaussian distributions in color-magnitude space, we select only the galaxies 99\% likely to be part of the blue cloud distribution as opposed to the red sequence distribution.  These selection criteria yield a total of 685 galaxies that are morphologically spiral and 1887 galaxies that are blue cloud members.  While there is overlap in these two populations (see Figure 2), the ``spiral" sample includes the red galaxies at the bright end of the blue cloud, while the ``blue cloud" sample includes fainter blue galaxies but not the red spirals at the bright end. (Over half of the faint blue cloud galaxies that do not make it into the spiral sample are preferentially identified as spiral or edge-on disk, just not at the 80\% confidence level.) The expected segregation of galaxies by type in different environments is evident in Figure 2 through the strong horizontal red sequence relative to the blue cloud in the inner 1.5 Mpc (left panel).  

\subsection{Upper limits on non-detections in HI} 

The limiting flux for galaxies not detected by ALFALFA can be estimated as 
$$S_{21,lim}= \sqrt{W_{50}\ 20}\  \sigma_{rms}\ S/N_{lim}$$ 
where $W_{50}$ is the expected 21-cm line width, the signal to noise $S/N_{lim}$ is set at 6.5, and the rms noise $\sigma_{rms}$ is measured directly for the location of each optically-detected galaxy not detected in HI (Gavazzi et al. 2013). 

We estimate W$_{50}$ following Springob et al. (2007): 
$$M_{I}=-7.85(\log W-2.5)-20.85+\log h.$$ 
The width $W$ must be corrected for inclination and broadening to give the expected observed width $W_{50}$.  
To correct for inclination, we multiply the width by $\sin i$.  The inclination angle $i$ is determined from the axial ratio according to 
$\cos 2i=((b/a)^{2}-q^{2})/(1-q^{2})$  where $b/a$ is the axial ratio from the SDSS and $q=0.20$ (see Springob et al. 2007). When $i < q$, $i$ is set to $90\degr$.  Finally, we add 6.5 km/s to account for broadening by internal motions, multiply by $1+z$ to account for broadening by redshift, and add 10 km/s to account for instrumental broadening, as described in Springob et al. (2007).

To measure $\sigma_{rms}$ for non-detections, we extract spectra from ALFALFA grids following the procedure described by Fabello et al. (2011). We integrate the grid over a region 4x4 arcmin$^{2}$ centered on the optical position of the galaxy and over all velocities. The $\sigma_{rms}$ for each polarization is determined outside of the estimated width of the galaxy and eliminating low weight regions $(<50\%)$. In this procedure, the width of each galaxy is estimated using the Tully-Fisher relation and the SDSS i-band magnitude following Giovanelli et al. (1997), adopting $\gamma=1.0$ in the internal extinction correction $\Delta m = - \gamma \log(a/b)$ (their equation 8), which is appropriate for all but the brightest galaxies and overcorrects for fainter galaxies (see Figure 7c of Giovanelli et al. 1995); this ensures that $\sigma_{rms}$ is measured outside any possible galaxy signal. 
We fit the baseline with a first-order polynomial and calculate $\sigma_{rms}$ about this fit, using the Arecibo IDL program robfit\_poly.pro by Phil Perrilat. We examine each galaxy spectrum after the fitting for the presence of RFI, baseline problems, and confusion and exclude galaxies from the sample if any of these is significant. Ninety galaxies are excluded, most because of RFI contamination. For the non-detections in our sample, the distribution of $\sigma_{rms}$ has a mean of 2.44 mJy and a standard deviation of 0.30 mJy.

\section{Results}

\subsection{Qualitative dependence of properties on group-centric distance and stellar mass}

Figures 3 and 4 illustrate the general behavior of galaxy properties as a function of group-centric distance for the blue cloud sample. 
The behavior of stellar mass $M_\star$ and galaxy diameter $D$ are shown in Figure 3. While both $\log M_\star$ and $\log D$ decrease slightly with group-centric distance, simple least squares linear regressions (black lines) yield a slope for $\log M_\star$ that is distinguished from zero at a much higher level of significance: $-0.114\pm 0.032$ for $\log M_\star$ as opposed to $-0.011\pm 0.011$ for $\log D$.  
The behavior of HI content is shown in Figure 4, with galaxies detected in HI on the left, and galaxies with HI detection limits on the right. For the HI detections on the left, all three measures of HI content vary with radius at the $95\%$ confidence level or better:  from top to bottom, the slopes are $0.066\pm 0.035$, $0.0142\pm 0.065$, and $-0.121\pm 0.031$. 
For the HI detection limits on the right, the large number of blue cloud galaxies in group centers that are not detected in HI appear as a concentration of limits at small $r/R_{200}$. 

A goal in this paper is to control for the dependence of HI properties on stellar mass.  As shown in Figure 5, HI mass and mass ratio depend strongly on stellar mass, while HI Deficiency increases slightly with stellar mass. 
Each of these slopes differs from zero at a confidence level greater than 99\%:  from top to bottom the slopes are $0.255\pm0.021$, $-0.745\pm0.023$, and $0.076\pm0.023$.
If stellar mass is not accounted for, HI properties will vary with group-centric distance simply because the stellar mass does.  In the analysis that follows, we account for stellar mass by limiting the range to $10.0<\log M_{\star}/M_{\sun}<10.5$ when comparing in-group galaxies to the control sample, and by incorporating stellar mass as an additional independent variable when running regressions. 
 
Figures 2-5 illustrate both the scatter in our sample and the approximately linear relationships among these quantities. 
This qualitative behavior motivates our use of fits to straight lines in the analysis that follows.
The regressions shown in these figures do not, however, take multiple variables into account, nor do they statistically combine the information from HI detections with HI detection limits.
In the analysis that follows, we use multiple linear regressions and take advantage of the branch of statistics known as survival analysis to incorporate detection limits.  

It is also important to note that the relationships we find among these quantities (both in this section and in the analysis that follows) do not take into account  completeness limits for the \textit{optically}-determined properties, and therefore do not provide the true, intrinsic slopes in these relationships.  A previous study that uses a much larger sample to estimate the true scaling relations among HI and optically-determined properties for isolated galaxies in the ALFALFA survey is Toribio et al. (2011), the study we use to calibrate our measure of HI Deficiency.  
By contrast, we use regressions here as means to look for trends significantly different from zero. In cases where we compare measured slopes with \textit{each other},
we consider whether these values could have different systematic offsets, for example by including galaxies at systematically different distances.

\begin{figure}
\includegraphics[width=270pt]{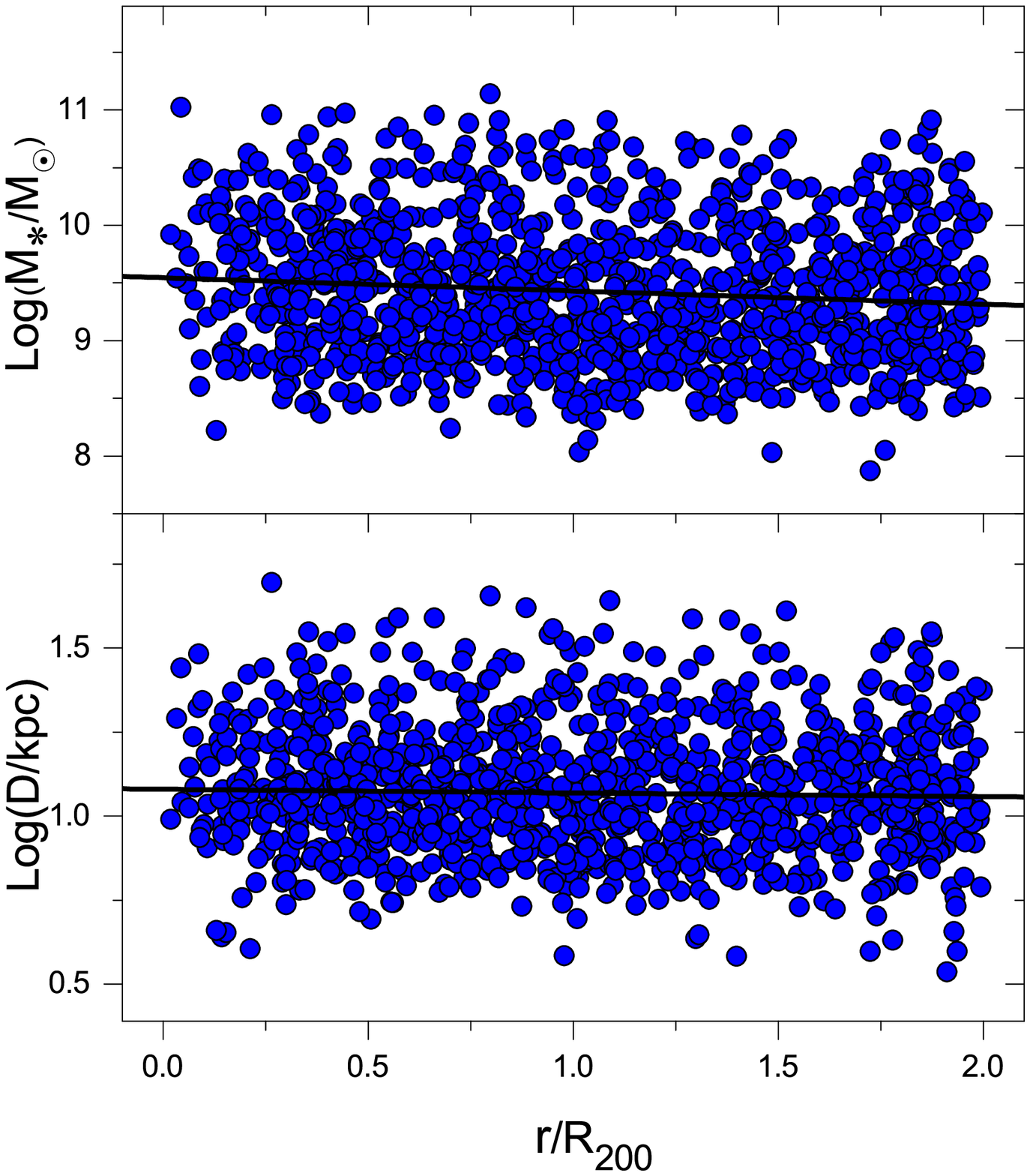}
\setlength{\abovecaptionskip}{-160pt} 
\caption{Dependence of stellar mass and galaxy diameter on the normalized group-centric distance for blue cloud galaxies. Solid lines show linear fits, which are described in the text.} 
\vspace*{-20mm}
\end{figure}
\begin{figure*}
\includegraphics[width=575pt]{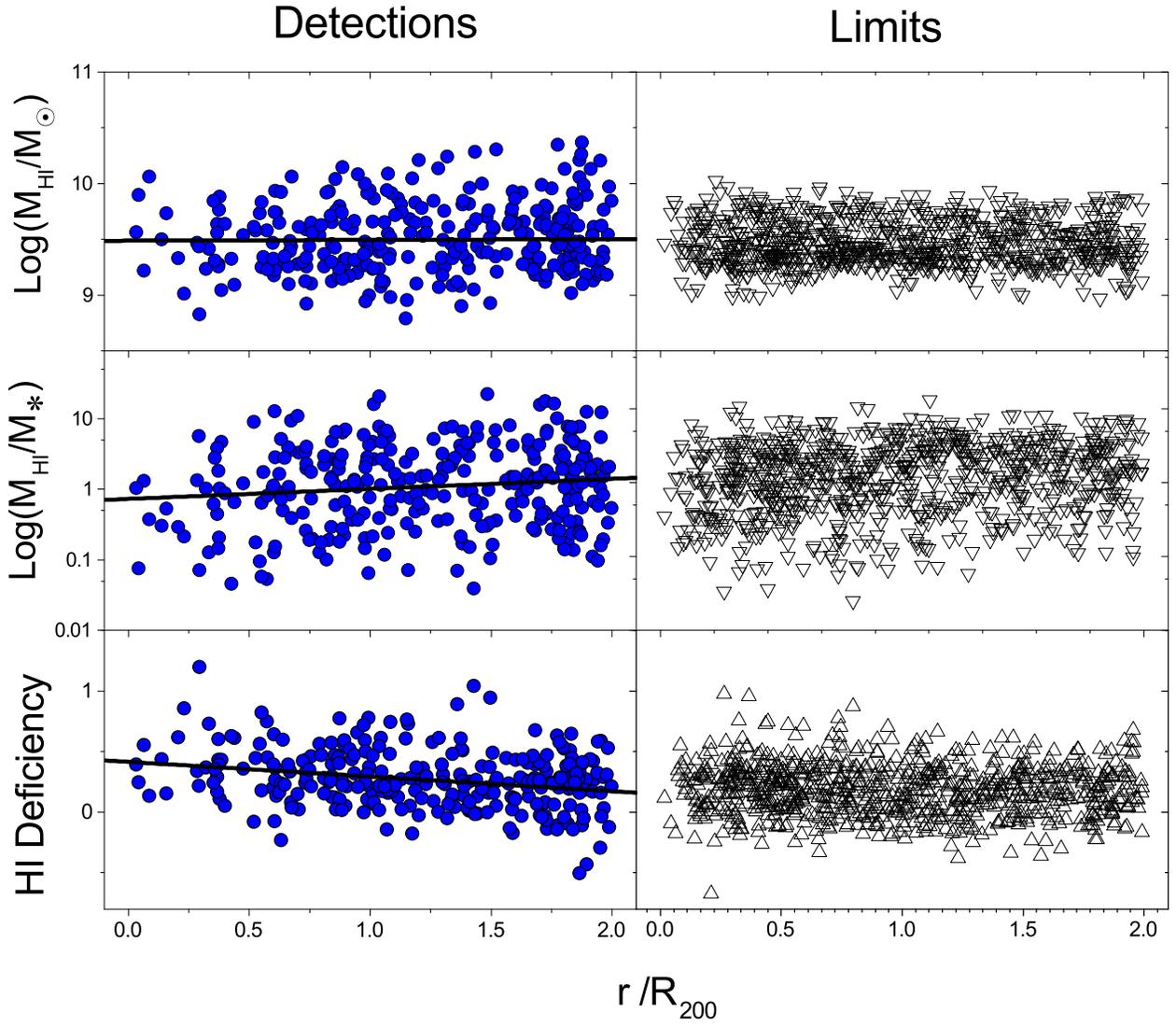}
\setlength{\abovecaptionskip}{-550pt} 
\caption{Dependence of HI mass, HI mass ratio, and HI Deficiency on the normalized group-centric distance for blue cloud galaxies. Detections in HI are shown as blue dots on the left, and limits are shown as open triangles on the right. Solid lines show linear fits to the HI detections. HI mass and mass ratio increase slightly, and HI Deficiency decreases slightly, with group-centric distance. The large number of late-type galaxies in group centers that are not detected in HI appear as a concentration of limits at small $r/R_{200}$ in the panels on the right.} 
\end{figure*}
\begin{figure}
\includegraphics[width=280pt]{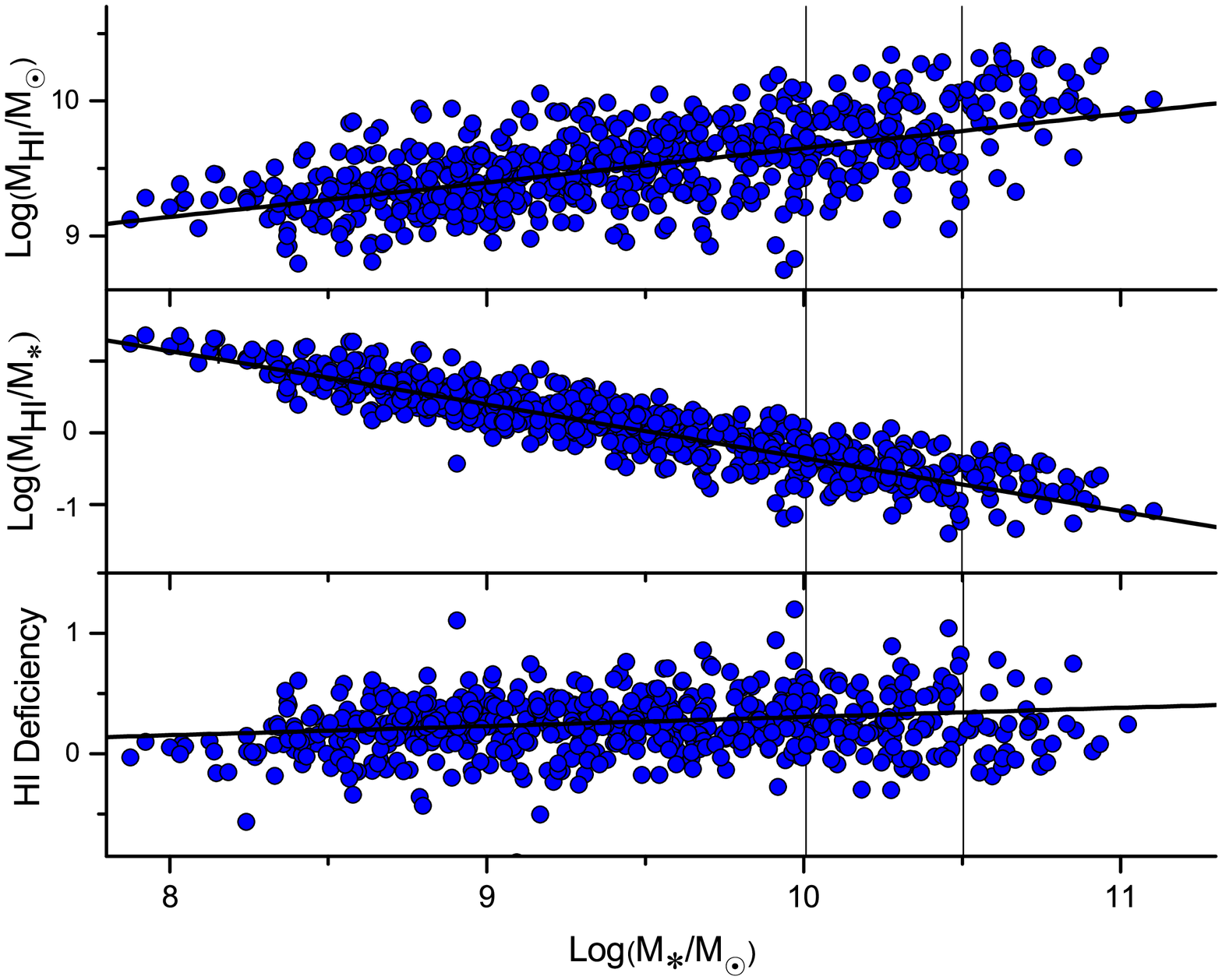}
\setlength{\abovecaptionskip}{0pt} 
\caption{The dependence of HI content on stellar mass for blue cloud galaxies.  Vertical lines indicate the limited range of stellar masses considered separately in Table 2. HI mass and mass ratio depend strongly on stellar mass, while HI Deficiency increases slightly with stellar mass.  If stellar mass is not accounted for, HI properties will vary with group-centric distance simply because the stellar mass does.}
\vspace*{5mm}
\end{figure}

\subsection{Is HI content at fixed $M_{\star}$ in groups different from the control sample?}

In this section, we first assess the differences between the HI content of late-type galaxies within groups (defined as a group-centric radius less than 1.5 Mpc) and the HI content of late-type galaxies in the control regions immediately surrounding them (group-centric distances 2.0-4.0 Mpc). 
Ideally, we would do this for each group or cluster individually, but many of these systems do not have enough galaxies to provide statistically significant results. Instead, we stack the groups by group-centric distance and compare the aggregate group and control samples.  
To address the possibility of pre-processing in groups before galaxies enter clusters, we focus on the groups in the lower two mass bins of Table 1. 

The HI content in group and control regions are compared using five standard statistics, three of which apply to detections only, and two of which are designed to include limits.  For detections only, we use the Kolmogorov-Smirnov test, the two-sided Anderson-Darling test, and the Wilcoxon rank-sum test. To include limits, we use the Log-rank and Peto-Peto tests, which compare so-called survival curves with different weightings. These statistical tests were run using the \textit{stats}, \textit{Matching}, \textit{kSamples}, and \textit{survival} packages in R. 

We find that galaxies in groups systematically lack HI relative to their control regions. Table 2 gives the probability values that the HI content of the group and control samples were drawn from the same population.  The five columns correspond to the five statistical tests, with the two tests that incorporate limits on the right. The table is split into four horizontal sections, based on whether the sample is chosen by spiral morphology, spiral morphology with limited stellar mass range, membership in the blue cloud, or membership in the blue cloud with limited stellar mass range. 

When the full range of stellar masses is included, both the spiral and blue cloud populations show statistically significant differences between the in-group and control samples, with consistently low p-values for both mass ratios and HI deficiencies. Differences in HI mass by itself 
are discernible only if limits are included.  When the range of stellar mass is limited to $10.0<\log M_{\star}/M_{\sun}<10.5$, the populations are less clearly distinguished (at least partly due to a much smaller sample size), but still yield p-values less than 0.05 in most cases. 
Note that differences in HI mass are \textit{better} distinguished in the limited stellar mass range, as might be expected from the very slight increase in HI mass with group-centric distance for the full sample of groups and clusters shown in Fig. 4 compared with the stronger decrease in stellar mass with group-centric distance shown in Fig. 3.

Another way to test for a dependence of galaxy properties on group-centric distance $r$
is through multiple linear regressions.  In this case, the dependence on stellar mass can be taken into account through regressions that include $\log M_{*}$ as an independent variable along with group-centric distance, rather than creating a separate mass-limited subsample. We perform the regressions with and without galaxy distance (assigned as the group distance given in Table 1) as an additional independent variable; the results below include group distance, but our qualitative conclusions are the same either way.  

As before, we perform the analysis on the detections as well as on the detections plus limits, using regression software designed to include limits using survival analysis statistics (in this case, the \textit{NADA} package in R).  
These results support the reality of the dependence of HI content on distance from group center, even at fixed stellar mass.  Table 3 shows
the probability that the slope is different from zero.  All the slopes determined using survival analysis are non-zero with high confidence.
In particular, the slopes for $\log M_{HI}/M_{\star}$ and HI Deficiency are distinguished from the zero with $p<0.001.$
Considering the limited sample of HI detections only, the HI Deficiency is still distinguished at the level of $p<0.01$. The significance of the slope for $\log M_{HI}/M_{\star}$ goes down considerably when $\log M_{\star}$ is included an additional variable. This is consistent with the behavior seen for the full sample of groups and clusters in Figs. 3-5:  
while $\log M_{HI}/M_{\star}$ increases with group-centric distance, part of this effect comes from having lower-mass galaxies (which tend to have higher values for $\log M_{HI}/M_{\star}$, at larger group-centric distance.

Note that when stellar mass is included as a separate variable, the p-values are the same for $\log M_{HI}$ and $\log M_{HI}/M_{\star}$. This is because, with $\log M_{\star}$ included as a separate variable in the regression, the remaining dependence of $\log M_{HI}/M_{\star}$ on group-centric distance is solely due to the dependence of $\log M_{HI}$ on group-centric distance.  

\begin{deluxetable*}{llllll}
\tablecolumns{6}
\tablewidth{0pc}
\tablecaption{Comparison of HI Content between Group and Control Samples}
\tablenum{2}
\tablehead{\colhead{}&\multicolumn{3}{c}{Detections only}& \multicolumn{2}{c}{Detections+limits} \\
\colhead{} & \colhead{K-S} & \colhead{A-D}   & \colhead{Wilcoxon}  & \colhead{Logrank} &\colhead{Peto-Peto} }
\startdata
\sidehead{Blue cloud galaxies}
\colhead{} & \multicolumn{3}{c}{$N_{in}=125, N_{out}=250$} & \multicolumn{2}{c}{$N_{in}=435, N_{out}=568$}\\
\\
Log($M_{HI}$)	 	&  0.1670     &  0.2756    & 0.2155    &  4.4e-4 ***   & 8.0e-4 ***   \\
Log($M_{HI}/M_{*}$)	&  6.5e-3 **  &  5.9e-4 ***& 5.3e-4 ***&  $<$1e-10 *** & $<$1e-10 ***   \\
HI Deficiency 		&  0.0921 .   &  0.0063 **& 0.0061 **&  5.3e-9 *** & 7.2e-10 ***   \\
\\
\sidehead{Blue cloud galaxies with $10.0<Log(M_{*})<10.5$}
\colhead{} & \multicolumn{3}{c}{$N_{in}=27, N_{out}=36$} & \multicolumn{2}{c}{$N_{in}=66, N_{out}=61$}\\
\\
Log($M_{HI}$)	  	&  0.0561 .   &  0.0496 *  & 0.0505 .   &  0.0319 * & 0.0086 **   \\
Log($M_{HI}/M_{*}$)	&  0.3950     &  0.0464 *  & 0.3031     &  0.0286 * & 0.0205 *    \\
HI Deficiency 		&  0.0970 .   &  0.0964 .  & 0.0648 .   &  0.0369 * & 0.0076 **   \\
\hline\sidehead{Spiral galaxies}
\colhead{} & \multicolumn{3}{c}{$N_{in}=81, N_{out}=123$} & \multicolumn{2}{c}{$N_{in}=178, N_{out}=216$}\\
\\
Log($M_{HI}$)	 	&  0.3762     &  0.4519  & 0.6210   &  0.0856 .  & 0.0569 .   \\
Log($M_{HI}/M_{*}$)	&  0.0397 *   &  0.0341 *& 0.0254 *&  5.2e-4 *** & 3.7e-4 ***   \\
HI Deficiency 		&  0.0362 *   &  0.0276 *& 0.0183 *&  3.9e-4 *** & 2.5e-4 ***   \\
\sidehead{Spiral galaxies with $10.0<Log(M_{*})<10.5$}
\colhead{} & \multicolumn{3}{c}{$N_{in}=26, N_{out}=31$} & \multicolumn{2}{c}{$N_{in}=53, N_{out}=51$}\\
\\
Log($M_{HI}$)	  	&  0.0122 *   &  0.0146 *  & 0.0119 *   &  0.2290  & 0.0430 *   \\
Log($M_{HI}/M_{*}$)	&  0.0183 *   &  0.0395 *  & 0.0358 *   &  0.1260  & 0.0432 *    \\
HI Deficiency 		&  0.0093 **  &  0.0218 *  & 0.0220 *   &  0.3160  & 0.0620    
\enddata
\tablecomments{P-values indicate the probability that group and control samples are drawn from the same population.  The five columns show results for each of five statistical tests. Three different measures of HI content are considered, for each of four ways of selecting the galaxy population: blue cloud galaxies, blue cloud within a limited mass range, spiral galaxies, and spiral galaxies within a limited mass range. The significance of p-values is denoted *** for $p<0.001$, ** for $p<0.01$, * for $p<0.05$, and . for $p<0.1$.}
\end{deluxetable*}

\begin{deluxetable*}{lrlrc}
\tablecolumns{5}
\tablecaption{Regression results for HI content as a function of group-centric distance} 
\tablenum{3}
\tablehead{\colhead{}&\multicolumn{2}{c}{Detections only}& \multicolumn{2}{c}{Detections+limits}  \\
\colhead{} & \colhead{slope}   & \colhead{p}  & \colhead{slope} &\colhead{p} }
\startdata
\sidehead{Blue cloud galaxies}
\colhead{} & \multicolumn{2}{c}{$N=423$} & \multicolumn{2}{c}{$N=1142$}\\
\\
Log($M_{HI})$	   & $-0.006\pm 0.012$& 0.6310       & $0.038 \pm 0.011$ &4.0e-4 ***\\
Log($M_{HI}/M_{*})$& $0.075\pm  0.023$ & 0.0011 .    & $0.138 \pm 0.020$  &$<$1e-10 ***\\
HI Deficiency      & $-0.035 \pm 0.011$ & 1.6e-3  ** & $-0.053 \pm 0.009$ &1.2e-9 ***\\
\sidehead{Blue cloud galaxies, including Log($M_{*}$)}\\
Log($M_{HI}$)	   & $0.017\pm 0.010$ & 0.0862 . & $0.048 \pm 0.008$  & 1.8e-8 ***\\
Log($M_{HI}/M_{*}$)& $0.017\pm 0.010$ & 0.0862 . & $0.048 \pm 0.008$  & 1.8e-8 ***\\
HI Deficiency      & $-0.032\pm 0.011$ & 3.8e-3 **& $-0.050 \pm 0.009$  & 8.1e-9 ***\\
\hline\sidehead{Spiral galaxies}
\colhead{} & \multicolumn{2}{c}{$N=230$} & \multicolumn{2}{c}{$N=441$}\\
Log($M_{HI})$	   & $0.008 \pm 0.016$ & 0.6334    & $0.037 \pm 0.016$ &0.0190 **\\
Log($M_{HI}/M_{*})$& $0.068 \pm 0.030$ & 0.0234 *  & $0.112 \pm 0.029$ &1.3e-4 ***\\
HI Deficiency      & $-0.044 \pm 0.015$ & 4.3e-3 ** &$-0.055 \pm 0.014$ &5.8e-5 ***\\
\sidehead{Spiral galaxies, including Log($M_{*}$)}\\
Log($M_{HI}$)	   & $0.023 \pm 0.014$ & 0.0999 . & $0.044 \pm 0.013$  & 7.7e-4 ***\\
Log($M_{HI}/M_{*}$)& $0.023 \pm 0.014$ & 0.0999 . & $0.044 \pm 0.013$  & 7.7e-4 ***\\
HI Deficiency      & $-0.040 \pm 0.015$ & 8.5e-3 **&$-0.052 \pm 0.014$  & 1.3e-4 ***
\enddata
\tablecomments{Slopes and the associated p-values indicating the probability that the slope is consistent with zero.  Columns represent results for linear regressions using HI detections only as well as linear regressions incorporating survivial analysis techniques to include HI detection limits.  As in Table 2, three different measures of HI content are considered, for both the blue cloud sample and the spiral sample. The significance of p-values is denoted *** for $p<0.001$, ** for $p<0.01$, * for $p<0.05$, and . for $p<0.1$.}
\end{deluxetable*}

The approaches summarized in Tables 2 and 3 support the conclusion that HI gas is deficient in groups, for fixed galaxy type and stellar mass, compared with control samples surrounding each group.  
This result holds true for late-type galaxies selected by either spiral morphology or position on the color-magnitude diagram.

These results are based on the groups in the two lower mass bins of Table 1.  Can we detect pre-processing if we limit the sample further, to only the lowest mass bin? 
In this case, we find statistically significant results for most, but not all, of the cases considered in Tables 2 and 3. Comparing the group and control samples as in Table 2, $\log M_{HI}/M_{\star}$ and HI Deficiency are different at the 
p$<$0.05 level for all cases when detection limits are included, and also for the full range of stellar mass (but not for the limited range) when only detections are included.  For the regressions of Table 3, $\log M_{HI}/M_{\star}$ and HI Deficiency yield slopes distinguishable from zero at the p$<$0.05 level for nearly all cases.  The only exception is that of $\log M_{HI}/M_{\star}$ when only detections are used and Log($M_{*}$) is included.  

How low can the group mass go and still show evidence of pre-processing at fixed stellar mass?  If we limit the sample further by systematically excluding the larger groups one by one, we find that $\log M_{HI}/M_{\star}$ and HI Deficiency yield slopes distinguishable from zero at the p$<$0.05 level down to a sample that includes the smallest six groups. (This is true only when non-detections are included, and when we consider the blue cloud sample rather than the smaller sample of spirals.) These six groups have velocity dispersions in the range $100-300$ km s$^{-1}$.

In the following sections, we consider the entire sample of groups and clusters to see how HI content depends on group or cluster mass, to compare different environmental predictors of HI content, and to examine how HI content depends on environment at both fixed stellar mass and fixed color.
These cases require splitting the sample into smaller subsets and/or adding additional variables to the regressions.  In each of these cases we focus on the blue cloud sample, and continue to control for stellar mass and group distance by including them as an independent variable in the regressions. 

\subsection{Environmental dependence of HI content as a function of group mass}

To further investigate the dependence of galaxy properties on group mass, we split the full sample into the four bins of Table 1, labeled according to their approximate masses, with $\log(M/M_\sun$) of 13.5, 14.0, 14.5, and 15.0.  (Coma is the only system in the largest mass bin.)  

We also consider X-ray luminosity as an alternate way of splitting the sample by size.  X-ray luminosity is expected to be related to both the  mass of the cluster and the amount of hot gas that could be directly responsible for stripping HI, and has been used as a measure of group size in previous studies of HI content. 
Sengupta \& Balasubramanyam (2006), for example, examined a sample that included ten X-ray detected groups over the range $40.5<L_{X}<43.0$ and eight groups with detection limits only.  They found that galaxies in the X-ray groups tended to be more HI Deficient than galaxies in the groups without X-ray detections. At the same time, they saw no systematic dependence of HI Deficiency on L$_{X}$. 

In order to look for differences beyond simple scaling with group radius, we use the normalized group-centric distance $r/R_{200}$ and fit only over the range $r/R_{200}<2$, a range sampled by all the groups.
Figure 6 shows slopes from regressions for stellar mass, g-i color, HI mass, and HI Deficiency.  Regressions for HI properties include limits, and both $\log M_{\star}$ and distance to the galaxy are included as additional independent variables.  (We do not include results for $\log M_{HI}/M_{\star}$ at fixed $M_{\star}$ since they give results identical to those for $\log M_{HI}$ at fixed $M_{\star}$, as discussed above.)  For ease of comparison, all slopes are shown as positive.
The directions of the slopes are consistent with those in the previous sections: HI mass increases with $r/R_{200}$, while stellar mass, g-i color, and HI Deficiency decrease with $r/R_{200}$. 

In the top panel of Figure 6, uncertainties in the slopes increase from small to large groups.  This is at least partly caused by the fact that there are more galaxies in the smaller mass bins.  The number of galaxies in each of the four mass bins is, from left to right) 673, 505, 415, and 300.  

In the bottom panel, we focus on the smaller groups in order to compare with the sample of Sengupta \& Balasubramanyam (2006).  All of these groups are within the lower two mass bins of the top panel. The number of galaxies is similar for each X-ray bin: from left to right, the numbers are 374, 341, and 322.  
Splitting the sample this way, and focusing only on smaller groups, we again see an increase in the magnitudes of slopes with group size, although considering the uncertainties in the slopes, the differences between the second two bins are not statistically significant.

We find a possible distinction between the slopes for HI content and those for stellar mass and color. The magnitudes of the slopes for HI mass and HI Deficiency systematically increase with group mass. The slopes for stellar mass and color, on the other hand, do not show a significant trend with group mass. 
This behavior suggests a scenario where HI gas is removed or cut off quickly in large groups and clusters, through an additional mechanism like ram-pressure stripping that is not effective in small groups.  

Although group distance is not perfectly uniform across the mass bins, we note that this is unlikely to be a source of the difference between small and large groups.  First, this difference remains when we include group distance as an independent variable in the regressions.  Second, the differences in average distance are small and do not vary systematically with group mass (see Table 1).  

\begin{figure}
\vspace{-0mm}
\includegraphics*[width=400pt]{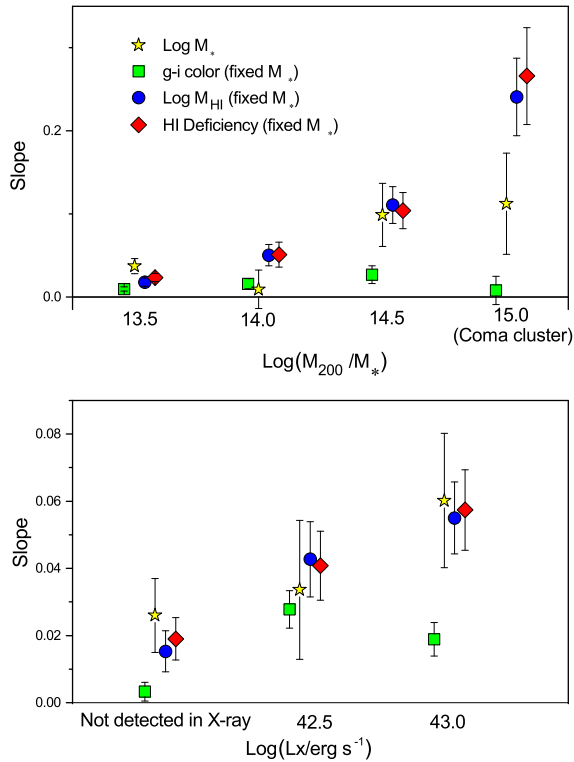}
\setlength{\abovecaptionskip}{-5pt} 
\caption{Slopes from regressions for log stellar mass, g-i color, log HI mass, and HI Deficiency for blue cloud galaxies, split by group mass (top panel) and by X-ray luminosity (bottom panel).  The top panel corresponds to the four group mass bins of Table 1. The bottom panel focuses on smaller systems only, showing a range of X-ray luminosities that spans only systems in the lower two mass bins. All regressions include stellar mass as an independent variable, and regressions for HI mass and HI Deficiency include detection limits. While behavior of stellar mass and color at fixed stellar mass do not exhibit a statistically significant trend with group size, the HI content does.}
\end{figure}

\subsection{Which environment variables are the best predictors of HI content?}

While the analysis above focuses on group-centric distance in order to look for pre-processing by comparing groups with the regions immediately surrounding them, HI content might be best described through another environment variable, or through multiple variables simultaneously.  It is clear from Figure 1, for example, that local density does not decrease smoothly with distance from the cluster center, suggesting that local density might be more sensitive description of environment, or that local density combined with group-centric distance might provide improved explanatory power.  

Here we present regressions against six environment variables to see which are the best predictors of galaxy properties:  group-centric distance $r$, normalized group-centric distance $r/R_{200}$, log density $\Sigma$, log group mass M$_{200}$ (as listed in Table 1), log halo mass in the Yang catalog, and central/satellite status in the Yang catalog.  In order to compare regressions for different physical quantities with each other, the slopes are standardized by dividing the values of each quantity by their standard deviation.  

A complication to expanding the range of variables beyond group-centric radius is that 
it becomes especially crucial to correct for the distance to each group, which ranges from 74 Mpc to 157 Mpc. Unlike group-centric distance, the other variables are not equally sampled for each group. As an example of a possible distance effect, the closer groups tend to sample more low-density environments, allowing us to see galaxies that are more HI deficient (because they are closer and easier to see) preferentially in low-density environments.  Stacking the sample by group-centric radius, as we do in the analysis above, deals with this by providing a control region for each group at the same distance as the group itself.  As in our other regressions, we correct for galaxy distance by including it as an additional independent variable.  Unlike in our previous regressions, however, our results are noticeably affected by this correction.

Figure 7 shows the results in the same format as Figure 6.  Local density $\Sigma$ is the strongest predictor of galaxy properties in the sense of having the highest standardized slopes and those most significantly distinguished from zero.  Both $r$ and $r/R_{200}$ are slightly less strong, followed by M$_{200}$ and M$_{halo}$, and lastly by central/satellite status. 
These last two categories have larger uncertainties in their slopes partly because they rely on the smaller sample that have halo properties listed in the Yang catalog (1045 out of the 1887 blue cloud galaxies).  Note also that the central/satellite distinction, being a binary categorical variable, is qualitatively different from the other variables. It is considered in more detail below.
Note also that, despite standardizing the slopes, the slopes and uncertainties for HI mass and HI Deficiency are systematically higher overall than those for stellar mass and color.  This is because we use the standard deviation for the detections alone, whereas the fitting process involves the limits as well.

To see whether fits are improved by combining multiple environment variables at once, we ran regressions on all combinations of two environment variables.  While this produced several cases in which both variables had slopes significantly different from zero, in very few cases was the fit significantly improved in terms of an adjusted $R^{2}$ or reduced $\chi{^2}$.  The exception was the addition of $\log M_{200}$ as a second variable to $r$, $r/R_{200}$, or $\Sigma$. 

\begin{figure}
\includegraphics[width=300pt]{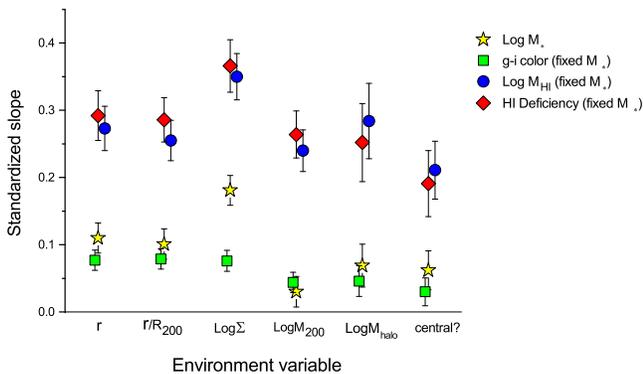}
\setlength{\abovecaptionskip}{-80pt} 
\caption{Standardized slopes from regressions for log stellar mass, g-i color, log HI mass, and HI Deficiency for blue cloud galaxies as a function of six different environment variables. All regressions include stellar mass as an independent variable, and regressions for HI mass and HI Deficiency include detection limits.
Slopes are all given as positive for ease of comparison.
Local density $\Sigma$ yields slopes that are the most clearly distinguished from zero.}
\end{figure}

\subsection{Comparison of central and satellite galaxies} 

An area of recent attention and controversy is the degree to which the quenching of star formation is different for central galaxies (typically defined as the brightest or most massive galaxy in a halo) and the satellite galaxies surrounding them (e.g. Peng et al. 2012, Tal et al. 2014, Knobel et al. 2015). 
Differences in the quenched fraction or quenching efficiency as a function of mass, local density, and redshift may imply different mechanisms for the quenching process, such as ``mass quenching" for centrals and ``environment quenching" for satellites. 
At the same time, as emphasized by Knobel et al. (2015), the results for these differences are sensitive to exactly how the populations are compared. In particular, satellites tend to be in higher-density environments, because large, dense haloes have many satellites but only one central, by definition.  Thus, satellite galaxies are different in systematic ways beyond their role as satellites.

Galaxy HI content has the potential to be a sensitive probe of these differences, since the removal or destruction of cool HI gas that precedes the quenching of star formation may take place at different rates and in different environments for the central and satellites populations.  Of the blue cloud galaxies within 4 Mpc of our group centers, 588 are identified as central and 457 as satellites in the Yang catalog. 
In order to look for systematic differences in the HI content for these two populations in different environments, we examine the dependence of HI Deficiency on local density, Yang halo mass, and stellar mass. Of particular interest is the dependence of HI Deficiency on local density at fixed stellar mass.
As in the preceding regressions, the group distance is included as an additional independent variable.  

Table 4 summarizes our results. Intercepts are included along with slopes, to provide information on the overall level of HI Deficiency as well as the strength of the slope.  
The slopes that are distinguished from zero at a confidence level of $p<0.05$ are the dependence on $\log \Sigma$ for both populations, the dependence on $\log M_{\star}$ for the centrals, and dependence on $\log M_{halo}$ for the satellites. Comparing the slopes for centrals and satellites with each other instead of with zero gives less definitive results, given the uncertainties in each slope.  In no cases are the slopes different by more than twice their combined uncertainties. 
Considering HI Deficiency at typical values of Log$\Sigma=1$,
 LogM$_{\star}=9.5$, and LogM$_{halo}=13$, we find that the HI Deficiency is slightly more for satellites at fixed Log$\Sigma$ or LogM$_{halo}$, and HI Deficiency is about the same for the centrals and satellites at fixed LogM$_{\star}$. 

To summarize, while central and satellite galaxies in the blue cloud exhibit a significant dependence of HI content on local density, only centrals show a strong dependence on stellar mass, and only satellites show a strong dependence on halo mass.  At fixed density or halo mass, the HI Deficiency for satellites is slightly higher.  
As emphasized by Knobel et al. (2015), comparing the central and satellite populations is complicated by the fact that these population differ in ways beyond their status as centrals and satellites.  For example, satellites are, on average, in regions of higher density.  In our case, this complication means that, although we account for local density as an independent variable in our regressions, nonlinearity in the behavior of HI Deficiency with respect to $\log\Sigma$, $\log M_{\star}$, and $\log M_{halo}$ could produce spurious differences between the populations, since the two populations have different distributions in these variables. (Note that for our sample we do not detect any significant nonlinearities in these relationships; see, for example, the bottom panel of Fig. 3.) In a larger sample, part of this dependence could be handled by comparing HI content for centrals relative to the satellites in their own haloes only.

\begin{deluxetable*}{lrlr}
\tablecolumns{4}
\tablecaption{Regressions for HI Deficiency in Central and Satellite Galaxies} 
\tablenum{4}
\tablehead{\colhead{}&\colhead{Slope}&\colhead{p}&\colhead{Intercept}} \\
\startdata
\sidehead{Centrals, N=588} \\

HI Deficiency(Log$\Sigma)$ &  $0.157\pm0.026$ & 1.3e-9*** & $0.755\pm0.081$ \\
HI Deficiency(LogM$_{halo})$& $-0.032\pm0.036$ & 0.374 & $1.178\pm0.469$ \\

\sidehead{Centrals, including Log$M_{\star}$} \\
HI Deficiency(Log$\Sigma)$ &  $0.144\pm0.025$& 1.3e-8*** & $0.227\pm0.203$ \\
HI Deficiency(LogM$_{halo})$& $-0.024\pm0.035$& 0.493 & $0.484\pm0.536$ \\
\\
HI Deficiency(LogM$_{\star}$)&$0.079\pm0.022$ &4.7e-4***&$0.512\pm0.210$ \\
\hline
\sidehead{Satellites, N=457} \\
HI Deficiency(Log$\Sigma)$ &  $0.174\pm0.035$& 6.1e-7*** &$0.979\pm0.125$ \\
HI Deficiency(LogM$_{halo})$& $0.097\pm0.026$& 2.1e-4*** &$-0.168\pm0.384$ \\

\sidehead{Satellites, including logM$_{\star}$} \\
HI Deficiency(Log$\Sigma)$ &  $0.174\pm0.036$& 2.5e-6*** & $0.768\pm0.343$ \\
HI Deficiency(LogM$_{halo})$& $0.091\pm0.026$& 4.8e-4*** & $-0.484\pm0.459$\\
\\
HI Deficiency(LogM$_{\star}$)&$0.059\pm0.034$& 0.087 . &$0.645\pm0.340$
\enddata
\tablecomments{Regressions for HI Deficiency as a function of local density Log$\Sigma$ and Yang halo mass LogM$_{halo}$, with and without LogM$_{\star}$ as an additional independent variable, for blue cloud galaxies that are centrals (top half) and satellites (bottom half). The final row in each half shows the dependence of HI Deficiency on LogM$_{\star}$ alone, without either Log$\Sigma$ or LogM$_{halo}$. Columns show slopes, p-values indicating the probability that the slope is consistent with zero, and y-intercepts. All results are for regressions that include survival analysis techniques to incorporate HI detection limits.  The significance of p-values is denoted *** for $p<0.001$, ** for $p<0.01$, * for $p<0.05$, and . for $p<0.1$.}
\end{deluxetable*}

\subsection{Environmental dependence of HI content as a function of galaxy mass}

Previous studies find that the quenching of star formation as a function of environment depends on galaxy mass --- for example, that quenched dwarf galaxies are found only in clusters (e.g. Baldry et al. 2006, Cortese et al. 2011, Geha et al. 2012), that fewer dwarf galaxies than expected from simulations are quenched (Wheeler et al. 2014), and that only in lower-mass galaxies does HI content have a stronger change with density than does specific star formation rate (Fabello et al. 2012).

In order to address differences in quenching with stellar mass, we split our population into high and low mass samples, each with 943 galaxies. As illustrated in Fig. 8, we find that regressions for HI mass, HI Deficiency, and color (all at fixed stellar mass) as function of group-centric distance yield slopes more strongly distinguished from zero for higher mass galaxies than for lower mass galaxies. Regressions for stellar mass, on the other hand, are more strongly distinguished from zero for the lower mass galaxies.  This would seem to imply that the star formation properties of dwarf galaxies are less sensitive to environment than are the star formation properties of larger galaxies, at least for the blue cloud.  
This could be interpreted as surprising, if the smaller galaxies are expected to be more sensitive to environment.
 
We interpret these results with special caution. The high-mass and low-mass samples have differences that are particularly tied to completeness in both HI and optical data. The cutoff in detection on the low-mass end could flatten slopes for the lower mass galaxies.  A full treatment including completeness would essentially be a calculation of the HI Deficiency function (analogous to the stellar mass function or HI mass function, e.g. Haynes et al. 2011) for different environments, which is beyond the scope of this paper. 

An additional complication is that smaller galaxies tend to appear in lower-density environments, so this trend may reflect the weaker environmental dependence for smaller groups illustrated in Figure 6. A larger sample would allow the effects of galaxy size and group size to be separated. 

\begin{figure}
\vspace{-5mm}
\includegraphics[width=550pt]{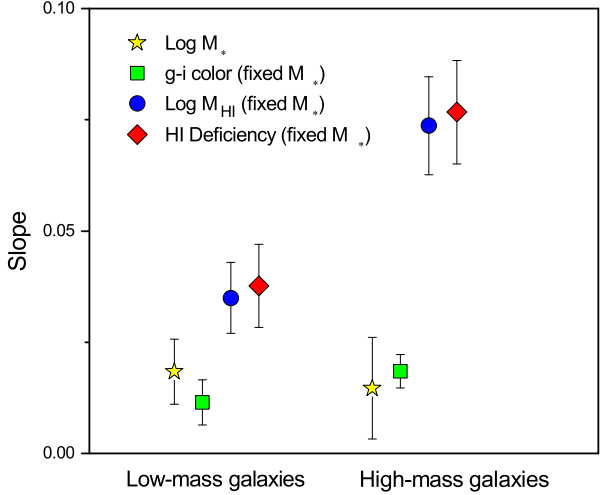}
\setlength{\abovecaptionskip}{-350pt} 
\caption{Slopes from regressions for log stellar mass, g-i color, log HI mass, and HI Deficiency for blue cloud galaxies as a function of group-centric distance, split into the 943 smaller galaxies and 943 larger galaxies, as measured by stellar mass. 
Slopes are all shown as positive for ease of comparison. All regressions include stellar mass and group distance as independent variables, and regressions for HI mass and HI Deficiency include detection limits. All properties except stellar mass show slopes more strongly distinguished from zero for the larger galaxies, but see the text for caveats.}
\vspace{-45mm}
\end{figure}

\subsection{HI Content for fixed stellar mass and color} 

If an environment-dependent depletion of HI gas is followed by reddening of the stellar population, we might see that the HI content of galaxies of fixed stellar mass \textit{and color} depends on environment.  Specifically, we can compare the HI content of galaxies in different environments that have the same stellar mass and color, and see whether the galaxies in denser environments lack HI gas.
To address this question, we compare regression results for HI Deficiency as a function of $\log M_{\star}$ plus $g-i$ color with regressions that include $\log\Sigma$ as well. As in the previous subsections, we consider the blue cloud sample and correct for group distance by including it as an independent variable.

We find that, indeed, the addition of local density increases the quality of fit as measured by the adjusted $R^{2}$, with HI Deficiency increasing with density. This dependence on environment at fixed stellar mass and color is strong when non-detections are included in the analysis (standardized slope $0.29 \pm 0.04$), but is also significant if only HI  detections are included (standardized slope $0.13 \pm 0.04$). 

The implication that galaxies of fixed color and mass are HI deficient in denser environments is supported by regressions for color as a function of stellar mass, HI, and local density.  
In this case, the addition of local density as an additional variable again improves the fit, but color becomes \textit{bluer} with increasing density.  This is the opposite of what is seen if HI Deficiency is not included as an independent variable, but it is what one would expect if a depletion of HI in denser environments precedes reddening within the blue cloud:  in order to match the HI Deficiency for a galaxy of the same mass in a less-dense environment, a galaxy in a denser environment must be bluer.

\subsection{How does the slope change with radius?}

Finally, we examine whether we can see a dependence on group-centric distance that is more complicated than a simple linear relationship.  While the general behavior of galaxy properties as a function of group-distance (Figures 3 and 4) appear approximately linear and do not support the choice of a particular curve that is more complicated than this, it is possible to show that the HI content varies more strongly in the inner regions.  

Figure 9 shows the standardized slopes from regressions as a function of group-centric distance for the inner and outer halves of the galaxy population, split so that the same number of galaxies is in each sample, providing the same statistical strength.  The halfway point is r = 2.06 Mpc, with 943 blue cloud galaxies in each population.

All four galaxy properties exhibit gradients that are statistically distinguished from zero for the inner half, but not for the outer half.  The blue cloud galaxies within the inner half show gradients indicating that galaxies at lower group-centric distances have higher stellar masses, redder colors, lower HI masses, and higher HI deficiencies. While the gradients for the outer half are not well distinguished from zero, the existence of some variation in this outer region is supported by the fact that the gradients for these properties are more strongly distinguished from zero for the entire sample than for the inner sample alone.  

The fact that the full sample includes a wide range of group and cluster sizes suggests that we might see a stronger signal if we stack the galaxies according to $r/R_{200}$ instead of $r$.  However, if we include only galaxies out to $2R_{200}$ (the maximum value available for all groups, so that outer regions are not dominated by the smaller groups), the number of galaxies drops from 1887 to 1140, and we are not able to distinguish a non-zero slope for either half.
Similarly, if we split the sample into smaller ranges of group mass, as in Section 4.2, we are not able to distinguish the slope from zero slope for either half.  

For our sample, the halfway point at $r=2.06$ is at or beyond $R_{200}$ (up to almost seven times beyond; see Table 1).  Therefore, it is not surprising that the gradient in galaxy properties is primarily within this radius. Ideally, with a larger sample, one could find a specific (not necessarily linear) function to describe the change in HI content within $R_{200}$ as function of group-centric distance.  So far, this has proven to be a challenge;  for example, Yoon \& Rosenberg (2015) do not detect a radial gradient in HI mass ratio within $R_{200}$ at all, using a slightly larger sample (but not including upper limits for galaxies not detected in HI). 

\begin{figure}
\vspace{-27mm}
\includegraphics[width=550pt]{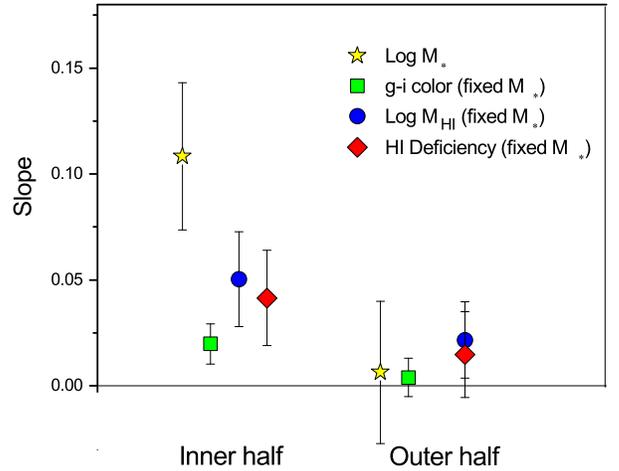}
\setlength{\abovecaptionskip}{-225pt} 
\caption{Slopes from regressions for log stellar mass, g-i color, log HI mass, and HI Deficiency for blue cloud galaxies as a function of group-centric distance, split into the 943 inner galaxies and 943 outer galaxies. 
Slopes are all shown as positive for ease of comparison. All regressions include stellar mass and group distance as independent variables, and regressions for HI mass and HI Deficiency include detection limits. Galaxies closer to group centers show a stronger gradient in properties. As discussed in the text, splitting the sample by group mass or fitting to $r/R_{200}$ did not yield slopes more distinguishable from zero.} 
\vspace{-23mm}
\end{figure}

\section{Conclusions and Discussion}

In this paper, we take the approach of comparing the HI content at fixed galaxy type and stellar mass in the centers of groups and clusters over a wide range of masses with the HI content in galaxies in control regions out to 4 Mpc surrounding each group or cluster. This process allows us to statistically test for pre-processing in the central systems themselves, including in isolated groups, an approach that complements environmental tests based on local density.  Our HI data is from the new $\alpha.70$ catalog, and we use statistical tests from survival analysis in order to include information on upper limits for galaxies not detected in HI.  

Our primary results are the following: 

1) Late-type galaxies in groups lack HI at fixed stellar mass relative to galaxies in a control region extending to 4.0 Mpc surrounding each group. Most of the gradient in galaxy properties out to 4.0 Mpc occurs in the inner 2.0 Mpc.  Although this effect is stronger in clusters and large groups ($M_{200}\gtrsim 10^{14.5}M_{\sun}$), it is also detectable in groups with $M_{200}\lesssim 10^{14.0}M_{\sun}$.  This shows pre-processing of HI gas, not simply at intermediate densities, but in intermediate-density isolated groups. 

2) The gradient in HI content for fixed stellar mass is stronger in larger groups, but the gradient in stellar mass and color does not follow this trend; stellar masses increase and galaxies become bluer with $r$ at similar rates across our four bins of group mass. These difference between HI and optical properties might reflect an additional, rapid mechanism for the removal of gas in the inner regions of larger groups and clusters.  Our statistics are insufficient to address whether this is truly a threshold effect, or a gradual change with group mass.

3) Comparing six different environment variables, we find that local density is the best predictor of galaxy properties (including HI content at fixed stellar mass), followed by group-centric radius $r$ and $r/R_{200}$. Considering all combinations of two variables together, we find that fits are improved (higher adjusted $R^{2}$) only for the addition of group mass $M_{200}$ to local density, $r$, or $r/R_{200}$.  We note that Papastergis et al. (2013) find that HI mass is not well predicted by halo mass, but may be predicted by halo spin.  The fact that we \textit{do} see a dependence on halo mass (with higher HI masses in smaller haloes), is at least partly due to the fact we compare galaxies at fixed stellar mass.  

4) While central and satellite galaxies in the blue cloud exhibit a significant dependence of HI content on local density, only centrals show a strong dependence on stellar mass, and only satellites show a strong dependence on halo mass.  At fixed density or halo mass, the HI Deficiency for satellites is slightly higher.  As emphasized by Knobel et al. (2015), comparing the central and satellite populations is complicated by the fact that these population differ in ways beyond their status as centrals and satellites.  For example, satellites are, on average, in regions of higher density.  In our case, this complication means that, although we account for local density as an independent variable in our linear regressions, non-linearity in the behavior of HI Deficiency with respect to $\log\Sigma$, $\log M_{\star}$, and $\log M_{halo}$ could produce spurious differences between the populations, since the two populations have different distributions in these variables. In a larger sample, part of this dependence could be handled by comparing the HI content for centrals relative to the satellites in their own haloes only.  

5) Splitting our sample into low-mass and high-mass galaxies, we find stronger gradients in high-mass galaxies for HI mass, HI Deficiency, and color (all at fixed stellar mass) as a  function of group-centric distance. On the other hand, we find stronger gradients in low-mass galaxies for stellar mass. While these results may imply that the star formation properties of dwarf galaxies within the blue cloud are less sensitive to environment than are the star formation properties of larger galaxies within the blue cloud, we caution that the high-mass and low-mass samples have differences that are particularly tied to completeness in both HI and optical data, and that these results do not take into account the dependence of galaxy stellar mass with group mass.  

6) We see evidence that HI is deficient for blue cloud galaxies in denser environments even when both stellar mass and color are fixed.  This is consistent with a picture where HI is removed or destroyed, followed by reddening within the blue cloud.

Our results support the existence of pre-processing in isolated groups, combined with a faster process of HI removal or destruction in larger clusters. The removal of HI gas is then followed by reddening within the blue cloud. More generally, our results support the picture that, even within the blue cloud and at fixed stellar mass, the HI and optical properties of galaxies vary with environment, a variation that happens faster or more recently than the full transition to a quiescent, red galaxy. 

\acknowledgements

This work was funded in part by National Science Foundation AST Award 1211005. MPH acknowledges support from NSF/AST-1107390 and the Brinson Foundation, and RAF acknowledges support from NSF/AST-0847430

Funding for the Sloan Digital Sky Survey IV has been provided by
the Alfred P. Sloan Foundation, the U.S. Department of Energy Office of
Science, and the Participating Institutions. SDSS-IV acknowledges
support and resources from the Center for High-Performance Computing at
the University of Utah. The SDSS web site is www.sdss.org.

SDSS-IV is managed by the Astrophysical Research Consortium for the 
Participating Institutions of the SDSS Collaboration including the 
Brazilian Participation Group, the Carnegie Institution for Science, 
Carnegie Mellon University, the Chilean Participation Group, the French Participation Group, Harvard-Smithsonian Center for Astrophysics, 
Instituto de Astrof\'isica de Canarias, The Johns Hopkins University, 
Kavli Institute for the Physics and Mathematics of the Universe (IPMU) / 
University of Tokyo, Lawrence Berkeley National Laboratory, 
Leibniz Institut f\"ur Astrophysik Potsdam (AIP),  
Max-Planck-Institut f\"ur Astronomie (MPIA Heidelberg), 
Max-Planck-Institut f\"ur Astrophysik (MPA Garching), 
Max-Planck-Institut f\"ur Extraterrestrische Physik (MPE), 
National Astronomical Observatory of China, New Mexico State University, 
New York University, University of Notre Dame, 
Observat\'ario Nacional / MCTI, The Ohio State University, 
Pennsylvania State University, Shanghai Astronomical Observatory, 
United Kingdom Participation Group,
Universidad Nacional Aut\'onoma de M\'exico, University of Arizona, 
University of Colorado Boulder, University of Oxford, University of Portsmouth, 
University of Utah, University of Virginia, University of Washington, University of Wisconsin, 
Vanderbilt University, and Yale University.

\textit{Facilities:} \facility{Arecibo}, \facility{Sloan}

\clearpage

\end{document}